\documentclass[aps,prl,nofootinbib,twocolumn,superscriptaddress]{revtex4-1}

\usepackage{epsfig}
\usepackage{color}
\usepackage[latin1]{inputenc}
\usepackage{float,amsmath, amstext, amssymb, amsfonts}
\usepackage{graphicx}
\usepackage{ulem}

\newcommand{\be}{\begin{eqnarray}}
\newcommand{\ee}{\end{eqnarray}}

\newcommand{\benum}{\begin{enumerate}}
\newcommand{\eenum}{\end{enumerate}}

\newcommand{\qs}{{Q_\mathrm{s}}}

\newcommand{\xt}{{\mathbf{x}_\perp}}
\newcommand{\yt}{{\mathbf{y}_\perp}}
\newcommand{\bt}{{\mathbf{b}_\perp}}

\begin{document}

\title{Fluctuating Glasma initial conditions and flow in heavy ion collisions}

\author{Bj\"orn Schenke}
\affiliation{Physics Department, Brookhaven National Laboratory, Upton, NY 11973, USA}

\author{Prithwish Tribedy}
\affiliation{Variable Energy Cyclotron Centre, 1/AF Bidhan Nagar, Kolkata 700064, India}

\author{Raju Venugopalan}
\affiliation{Physics Department, Brookhaven National Laboratory, Upton, NY 11973, USA}

\begin{abstract}
We compute initial conditions in heavy-ion collisions within the Color Glass Condensate (CGC) framework by
combining the impact parameter dependent saturation model (IP-Sat) with the classical Yang-Mills description of initial Glasma fields. In addition to fluctuations of nucleon positions, this IP-Glasma description includes 
quantum fluctuations of color charges on the length-scale determined by the inverse nuclear saturation scale $Q_s$.
The model naturally produces initial energy fluctuations that are described by a negative binomial distribution.
The ratio of triangularity to eccentricity $\varepsilon_3/\varepsilon_2$ is close to that in a model tuned to reproduce experimental flow data. We compare transverse momentum spectra and $v_{2,3,4}(p_T)$ of pions from different models of initial conditions
using relativistic viscous hydrodynamic evolution. 
\end{abstract}

\maketitle


A large uncertainty in the hydrodynamical description of ultrarelativistic heavy ion collisions is our imperfect knowledge of multi-gluon states in the nuclear wavefunctions and the early-time dynamics of gluon fields after the collision. In heavy ion collisions, studies of observables sensitive to harmonics of hydrodynamic flow distributions provide constraints both on the low shear viscosity of the Quark-Gluon Plasma (QGP) and the initial state dynamics~\cite{Alver:2010gr,Alver:2010dn,Petersen:2010cw,Schenke:2010rr,Schenke:2011tv,Qiu:2011iv,*Qiu:2011hf,Bleicher:2011ne,Schenke:2011bn}. This situation is analogous to studies of the cosmic microwave background \cite{Komatsu:2008hk}, wherein inhomogeneities in the observed power spectrum are  sensitive to primordial quantum fluctuations.

An {\it  ab initio} framework for multi-particle production is the Color Glass Condensate (CGC)~\cite{Gelis:2010nm} wherein the initial state dynamics is described by flowing Glasma gluon fields~\cite{Kovner:1995ts,*Kovner:1995ja,Krasnitz:1999wc,*Krasnitz:2000gz}.  There are several sources of quantum fluctuations that can influence hydrodynamic flow on an event-by-event basis. An important source of fluctuations, generic to all models of quantum fluctuations, are fluctuations in the distributions of nucleons in the nuclear wavefunctions. In addition there are fluctuations in the color charge distributions inside a nucleon.
This, combined with Lorentz contraction,  results in ``lumpy'' transverse projections of color charge configurations that vary event to event. The scale of this lumpiness is given on average by the nuclear saturation scale $Q_s$ which corresponds to distance scales smaller than the nucleon size~\cite{Kowalski:2007rw}.  For each such  configuration of color charges, the Quantum Chromo-Dynamics (QCD) parton model predicts dynamical event-by-event fluctuations in the multiplicities, the impact parameters and the rapidities of produced gluons~\cite{Miettinen:1978jb}. 

All these sources of fluctuations  are captured in the CGC Glasma flux tube picture. The relevant feature of this scenario is that long range rapidity correlations from the initial state wavefunctions, coherent over $1/Q_s$ in the transverse plane, are efficiently transmitted into hydrodynamic flow of the final state quark-gluon matter~\cite{Voloshin:2003ud,Shuryak:2007fu}. 

Recently, Monte-Carlo Glauber-type models (MC-Glauber) and Monte-Carlo implementations of the Kharzeev-Levin-Nardi-model (MC-KLN)~\cite{Hirano:2005xf,Drescher:2006pi,*Drescher:2007ax} have been compared to experimental data on elliptic and triangular moments of the flow distribution. While both types of models treat fluctuations in nucleon positions identically, the Glauber model implementations do not specify a mechanism for multi-particle production which would constrain the initial energy density distribution. MC-Glauber initial conditions~\cite{Esumi:2011nd,Qiu:2011iv,*Qiu:2011hf} can be tuned to reproduce data on both elliptic and triangular flow from RHIC and the LHC. The MC-KLN model is motivated by the CGC with approximations that will be discussed further below. It requires larger values of the viscosity to entropy density ratio ($\eta/s$) relative to the Glauber model values to describe elliptic flow data. This however leads it to underpredict triangular flow data.

Odd flow harmonics are entirely driven by fluctuations; it is therefore essential to have a realistic description of quantum fluctuations in multi-particle production to properly describe the final state dynamics.
Towards this end,  we will consider in this letter the impact parameter dependent saturation model (IP-Sat)~\cite{Bartels:2002cj,Kowalski:2003hm} to determine fluctuating configurations of color charges in two incoming highly energetic nuclei. This model is formally similar to the classical CGC McLerran-Venugopalan (MV) model of nuclear wavefunctions~\cite{McLerran:1994ni,*McLerran:1994ka,*McLerran:1994vd}, but additionally includes Bjorken $x$ and impact parameter dependence through eikonalized gluon distributions of the proton that are constrained\footnote{The IP-Sat model gives good $\chi$-squared fits to available small $x$ HERA data~\cite{Kowalski:2006hc} and fixed target e+A DIS data~\cite{Kowalski:2007rw}. Since the analysis of Ref.~\cite{Kowalski:2006hc}, more precise data is now available; a repeat analysis is desirable.}  by HERA inclusive and diffractive e+p deeply inelastic scattering (DIS) data~\cite{Kowalski:2006hc}. Most importantly, the model is in excellent agreement with data on n-particle multiplicity distributions in p+p collisions at RHIC and the LHC  and in A+A collisions at RHIC~\cite{Tribedy:2010ab,*Tribedy:2011aa}, an essential requirement for microscopic models. The MC-KLN model does not contain these features; a scheme to introduce fluctuations in the model has only been discussed recently~\cite{Dumitru:2012yr}.

The color charges $\rho^a(x^-,\xt)$ in the IP-Sat/MV model act as local sources for small $x$ classical gluon Glasma fields. These are determined by solving the classical Yang-Mills (CYM) equations $[D_{\mu},F^{\mu\nu}] = J^\nu$, with the  color current $J^\mu = \delta^{\mu \pm}\rho_{A(B)}(x^\mp,\xt)$ generated by a nucleus A (B) moving along the $x^+$ ($x^-$) direction.\footnote{Light cone quantities are defined as $v^\pm=(v^0\pm v^3)/\sqrt{2}$. The $\tau,\eta$ coordinates are defined as
$\tau = \sqrt{2 x^+ x^-}$ and $\eta = 0.5\ln(x^+/x^-)$. }
The solution in light cone gauge $A^+ (A^-) = 0$ are the pure gauge fields ~\cite{McLerran:1994ni,*McLerran:1994ka,*McLerran:1994vd,JalilianMarian:1996xn,Kovchegov:1996ty}   
 $A^i_{A (B)}(\xt) = \frac{i}{g}V_{A (B)}(\xt)\partial_i V^\dag_{A (B)}(\xt)
$ and $A^- (A^+) =0$. 
Here  $V_{A (B)} (\xt) = P \exp({-ig\int dx^{-} \frac{\rho^{A (B)}(x^-,\xt)}{\boldsymbol{\nabla}_T^2+m^2} })$ is a path ordered Wilson line in the fundamental representation, where the infrared regulator $m$ (of order $\Lambda_{\rm QCD}$) incorporates color confinement at the nucleon level.

The initial condition for a heavy-ion collision at time $\tau=0$ is given by the solution of the CYM equations in Schwinger gauge 
$A^\tau=0$, a natural choice because it interpolates between the light cone gauge conditions of the incoming nuclei. It has a simple expression in terms of the gauge fields of the colliding nuclei~\cite{Kovner:1995ja,Kovner:1995ts}:
\begin{equation}
\label{eq:init}
  A^i = A^i_{(A)} + A^i_{(B)}\,\,;\,\,A^\eta = \frac{ig}{2}\left[A^i_{(A)},A^i_{(B)}\right]\,,
\end{equation}
and\footnote{The metric in the $(\tau,\xt,\eta)$ coordinate system is $g_{\mu\nu} = {\rm diag}(1,-1,-1,-\tau^2)$ so that $A_\eta=-\tau^2 A^\eta$. The $\pm$ components of the gauge field are related by $A^\pm = \pm x^\pm A^\eta$.} $\partial_\tau A^i = 0$, $\partial_\tau A^\eta =0$. 
In the limit $\tau\rightarrow 0$, $A^\eta=-E_\eta/2$, with $E_\eta$ the longitudinal component of the electric field. At $\tau=0$, 
the only non-zero components of the field strength tensor are the longitudinal magnetic and electric fields, which can be computed non-perturbatively. 
They determine the energy density of the Glasma at $\tau=0$ 
at each transverse position in a single event~\cite{Krasnitz:1999wc,*Krasnitz:2000gz,Lappi:2003bi}.
The Glasma distribution computed from the CYM equations\footnote{As noted previously~\cite{Blaizot:2010kh}, the CYM approach treats soft modes with $k_\perp \leq Q_s$ more accurately than in commonly used $k_\perp$ factorized descriptions.}  (IP-Glasma) is matched event-by-event to viscous relativistic hydrodynamics~\cite{Schenke:2010rr,Schenke:2011tv} to compute harmonics of the flow distributions. 

We will now discuss details of the computation. Nucleon positions in the nucleus are sampled from a
Fermi distribution.
The saturation scale $Q_{s, (p)}^2(x,\bt)$ is determined from the IP-Sat dipole cross section for each nucleon, where $\bt$ is the impact parameter relative to each nucleon's center.  The  color charge squared per unit transverse area $g^2\mu^2(x,\bt)$ is proportional\footnote{The exact numerical factor between the two quantities depends on the details of the calculation \cite{Lappi:2007ku} but will not be relevant for our final results.} to $Q_{s, (p)}^2(x,\bt)$. For the nucleus, 
$g^2 \mu_A^2 (x,\xt)$
is obtained~\cite{Krasnitz:2002mn} by adding the individual nucleon $g^2\mu^2$ at the same $x$ and transverse position $\xt$ in the nucleus. 
\begin{figure}[h]
\centerline{\includegraphics[width=6.5cm]{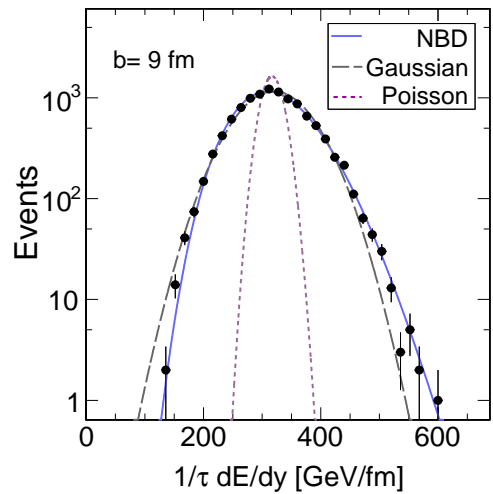}}\vspace{-0.2cm}
\caption{The IP-Glasma event-by-event distribution in energy for $b=9$ fm on the lattice compared to different functional forms. The negative 
binomial distribution (NBD) gives the best fit. 
\label{fig:tote}}
\vspace{-0.3cm}
\end{figure}

The lattice formulation of the Glasma initial conditions in Eq.~(\ref{eq:init}) was first given in \cite{Krasnitz:1998ns}. On a transverse lattice, random color charges\footnote{Here, and henceforth, the distributions are evaluated at 
$x= \langle p_\perp\rangle/\sqrt{s}$, for zero rapidity, where $\langle p_\perp\rangle$ is the average transverse momentum of charged hadrons in p+p collisions at a given $\sqrt{s}$.} $\rho^a(\xt)$ are sampled from
\begin{equation}
  \langle \rho_k^a(\xt)\rho_l^b(\yt)\rangle = \delta^{ab}\delta^{kl}\delta^2(\xt-\yt)\frac{g^2\mu_A^2(\xt)}{N_y}\,,
\end{equation}
where the indices\footnote{$N_y=100$ in all computations presented here.} $k,l=1,2,\dots,N_y$ represent a discretized $x^-$ coordinate~\cite{Lappi:2007ku}. For the large nuclei we consider the use of such local Gaussian color charge distributions is a valid approximation\footnote{Modifications to Gaussian distributions, relevant for smaller nuclei, have recently been explored in \cite{Dumitru:2011zz}.}.
 The path ordered Wilson line is discretized as 
\begin{equation}
  V_{A (B)}(\xt) = \prod_{k=1}^{N_y}\exp\left(-ig\frac{\rho_k^{A (B)}(\xt)}{\boldsymbol{\nabla}_T^2+m^2}\right)\,.
\end{equation}
To each lattice site $j$ we assign two SU($N_c$) matrices $V_{(A),j}$ and $V_{(B),j}$, each of which defines a pure gauge configuration with the link variables
$
  U^{i}_{(A,B),j} = V_{(A,B),j}V^\dag_{(A,B),j+\hat{e_i}}\,,
$
where $+\hat{e_i}$ indicates a shift from $j$ by one lattice site in the $i=1,2$ transverse direction. The link variables in the future lightcone $U_{j}^{i}$, are 
determined from solutions of the lattice CYM equations at $\tau=0$,
\begin{align}
  &{\rm tr} \left\{ t^a \left[\left(U^{i}_{(A)}+U^{i}_{(B)}\right)(1+U^{i\dag})\right.\right.\nonumber\\
   &  ~~~~~~~~~ \left.\left.-(1+U^{i})\left(U^{i\dag}_{(A)}+U^{i\dag}_{(B)}\right)\right]\right\}=0\,,\label{eq:initU}
\end{align}
where $t^a$ are the generators of $SU(N_c)$ in the fundamental representation (The cell index $j$ is omitted here).
The $N_c^2-1$ equations\,(\ref{eq:initU}) are highly non-linear and for $N_c=3$ are solved iteratively. 

The total energy density on the lattice at $\tau=0$ is given by
\begin{equation}
  \varepsilon(\tau=0) = \frac{2}{g^2 a^4} (N_c - {\rm Re}\,{\rm tr}\, U_\square) + \frac{1}{g^2 a^4} {\rm tr}\,E_\eta^2\,,
\end{equation}
where the first term is the longitudinal magnetic energy, with the plaquette given by
$  U_\square^j = U^x_j \, U^y_{j+\hat{x}} \, U^{x\dag}_{j+\hat{y}} \, U^{y\dag}_j$. The explicit lattice expression for the longitudinal electric field in the 
second term can be found in Refs.~\cite{Krasnitz:1998ns,Romatschke:2006nk}.  We note that the boost-invariant CYM framework neglects fluctuations in the rapidity direction.  Anisotropic flow at mid-rapdity is dominated by fluctuations in the transverse plane but fluctuations in rapidity could have an effect on the dissipative evolution; the framework to describe these effects has been developed~\cite{Dusling:2011rz} and will be addressed in future work. Other rapidity dependent initial conditions are discussed in Ref.~\cite{Magas:2000jx,*Magas:2002ge}.

In Fig.~\ref{fig:tote} we show the event-by-event fluctuation in the initial energy per unit rapidity. The mean was adjusted to reproduce particle multiplicities after hydrodynamic evolution. This and all following 
results are for Au+Au collisions at RHIC energies ($\sqrt{s}=200\, A\,{\rm GeV}$) at midrapidity. 
The best fit is given by a negative binomial (NBD) distribution, as predicted in the Glasma flux tube framework~\cite{Gelis:2009wh}; our result adds further confirmation to a previous non-perturbative study~\cite{Lappi:2009xa}. 
The fact that the Glasma NBD distribution fits p+p multiplicity distributions over RHIC and LHC energies~\cite{Tribedy:2010ab,*Tribedy:2011aa} lends confidence that our picture includes fluctuations properly.

\begin{figure}[tb]
   \begin{center}
     \includegraphics[width=4cm, clip, angle=270,trim=3.5cm 0cm 0cm 1cm]{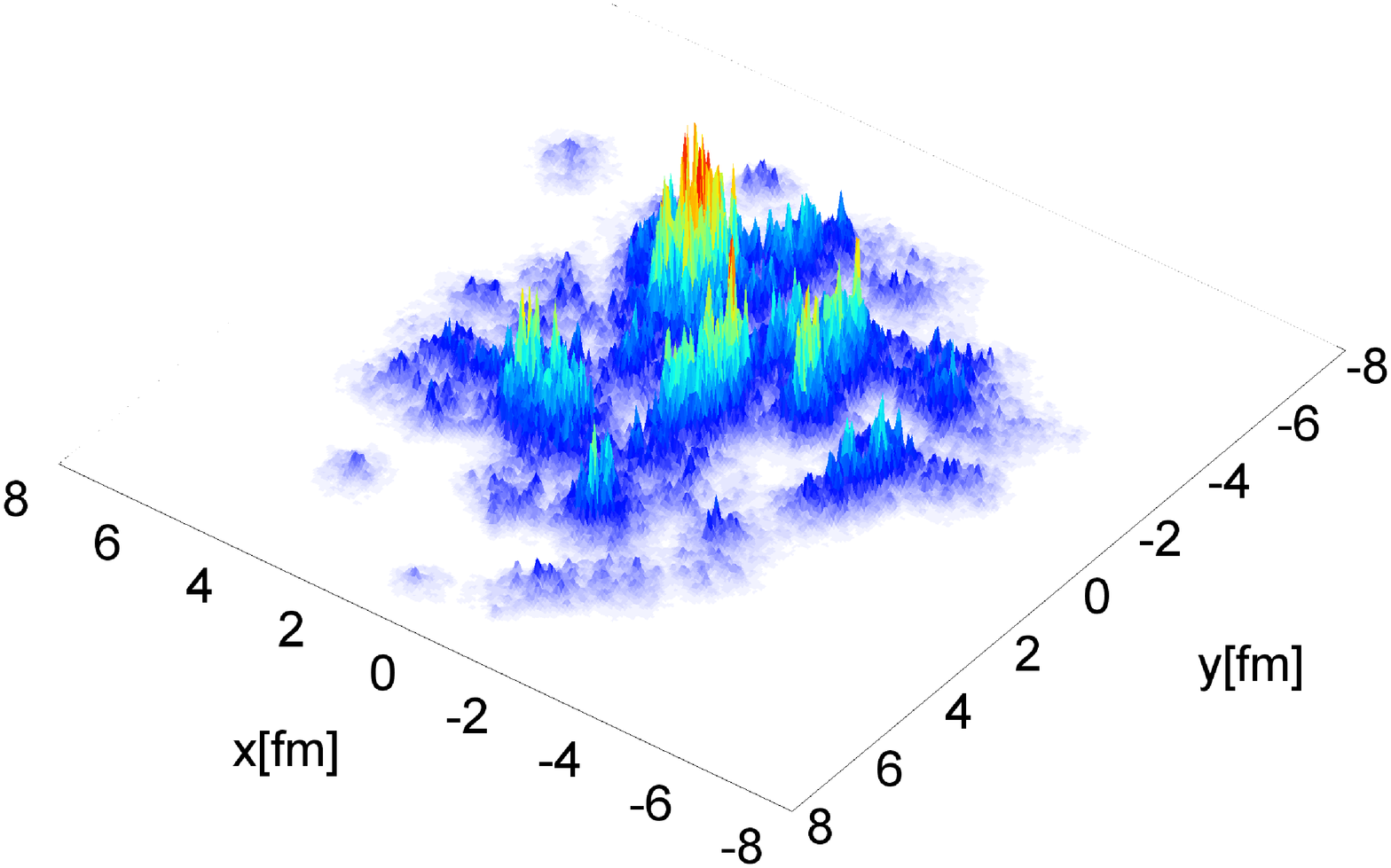}
     \includegraphics[width=4cm, clip, angle=270,trim=3.5cm 0cm 0cm 1cm]{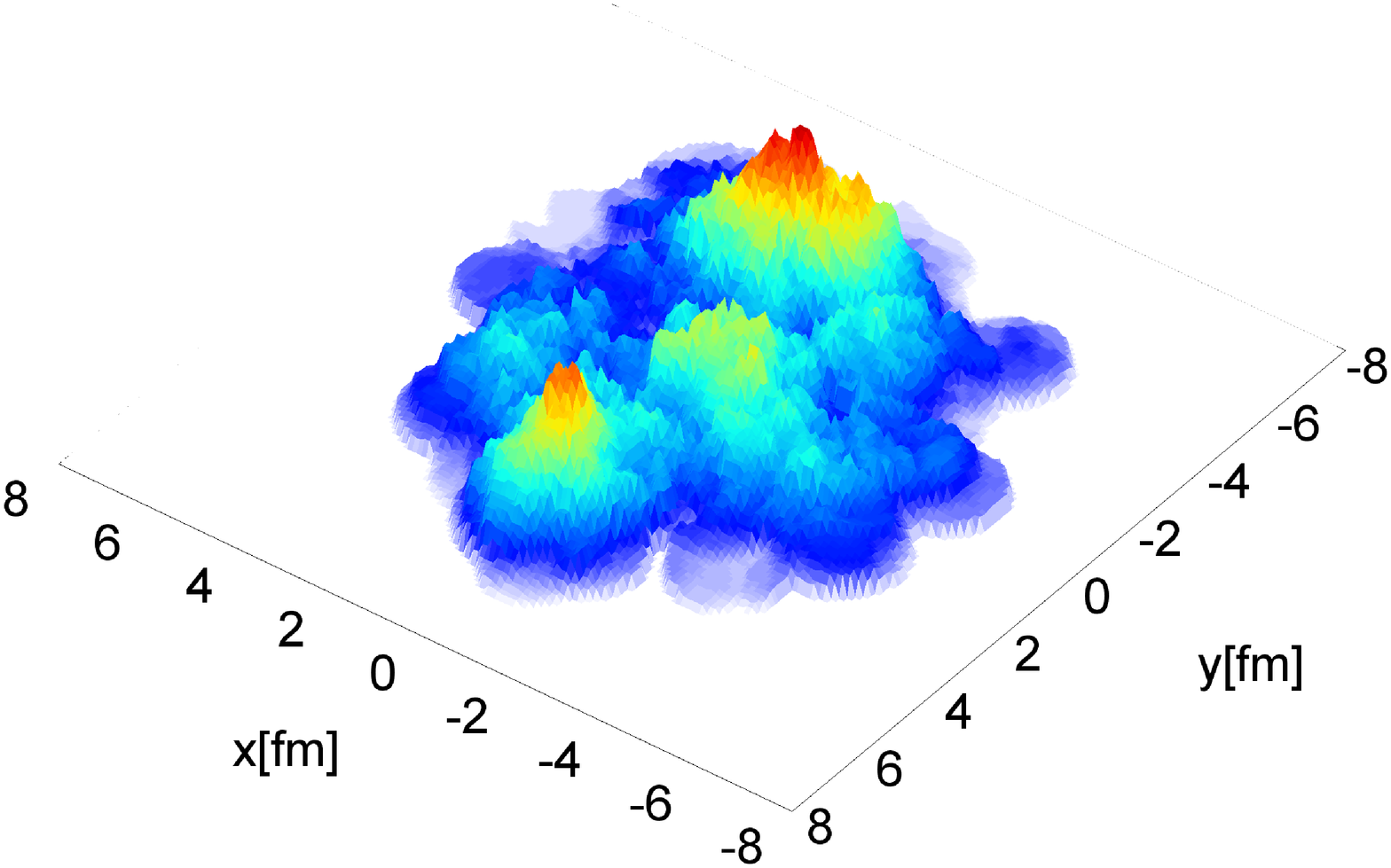}
     \includegraphics[width=4cm, clip, angle=270,trim=3.5cm 0cm 0cm 1cm]{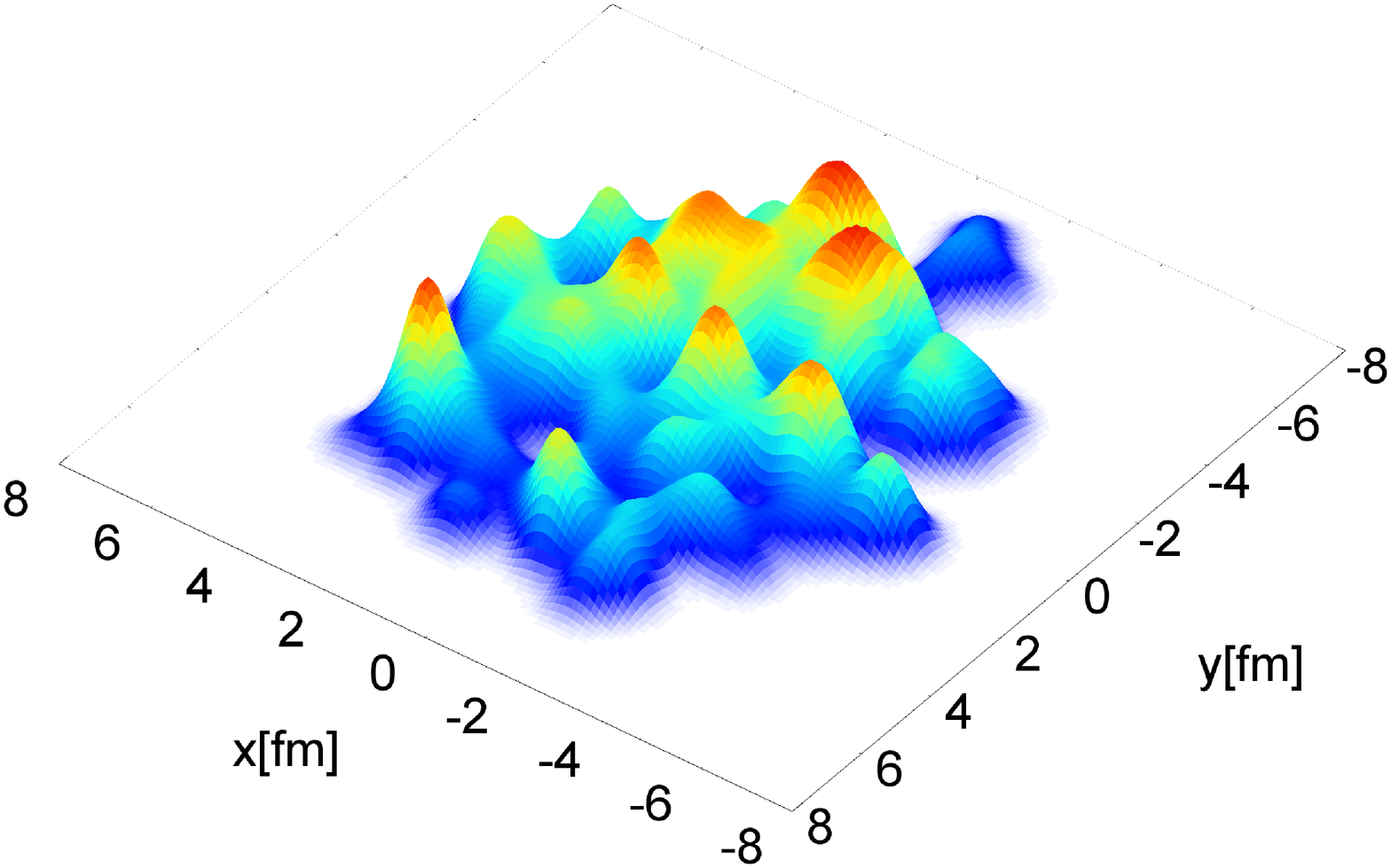}
     \vspace{-0.5cm}
     \caption{(Color online) Initial energy density (arbitrary units) in the transverse plane in three different heavy-ion collision events: from top to bottom, IP-Glasma,  MC-KLN  and MC-Glauber~\cite{Schenke:2011bn} models. }
     \label{fig:eps}
   \end{center}
\vspace{-0.7cm}
\end{figure}

We now show the energy density distribution in the transverse plane in Fig.~\ref{fig:eps}. 
We compare to the MC-KLN model and to an MC-Glauber model that was tuned to reproduce experimental data \cite{Schenke:2010rr,Schenke:2011bn}. 
In the latter, for every participant nucleon, a Gaussian distributed energy density is added.
Its parameters are the same for every nucleon in every event, with the width chosen to be $0.4\,{\rm fm}$ to best describe anisotropic flow data.
We will also present results for a model where the same Gaussians are assigned to each binary collision. 
The resulting initial energy densities differ significantly. In particular, fluctuations in the IP-Glasma 
occur on the length-scale $Q_s^{-1}(\xt)$, leading to finer structures in the initial energy density relative to the other models. 
 As noted in~\cite{Dumitru:2012yr}, this feature of CGC physics is missing in the MC-KLN model.

\begin{figure}[tb]
   \begin{center}
     \includegraphics[width=8cm, clip, trim=0cm 0.1cm 0cm 0cm]{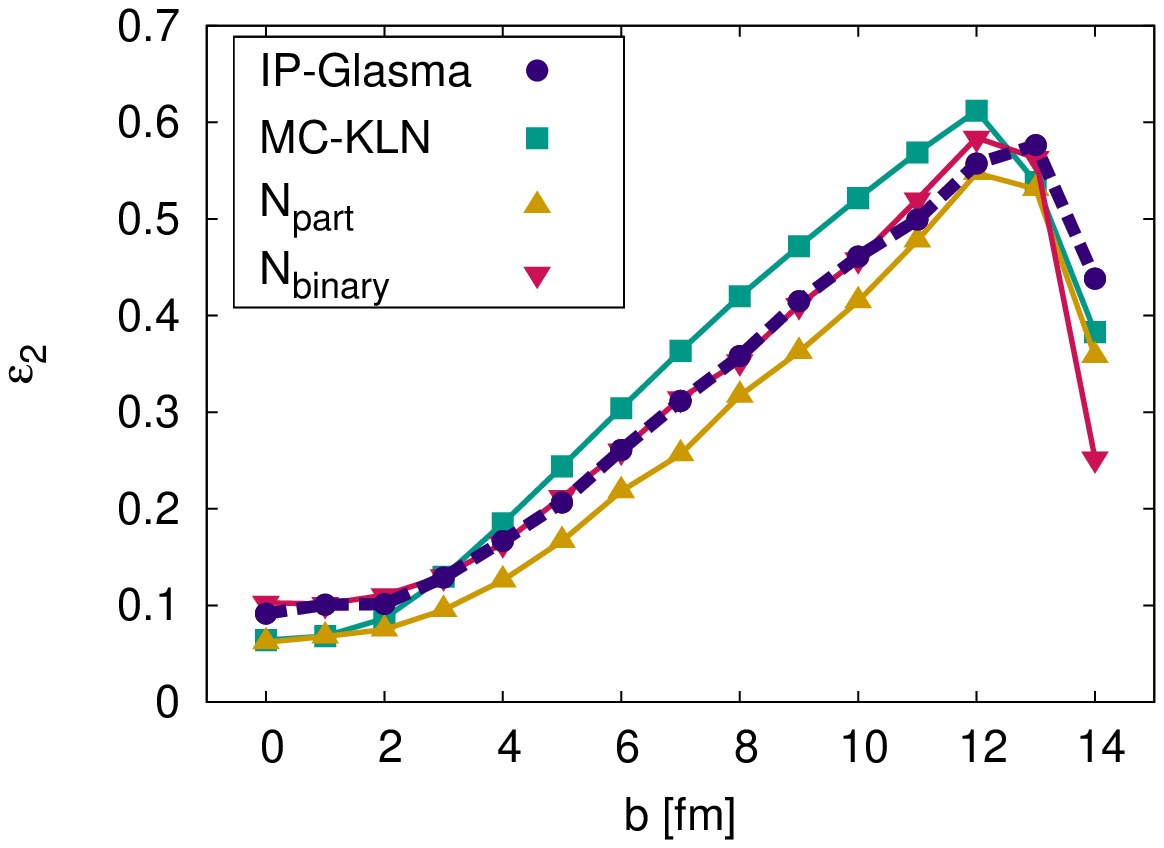}
     \includegraphics[width=8cm, clip, trim=0cm 0cm 0cm 0.2cm]{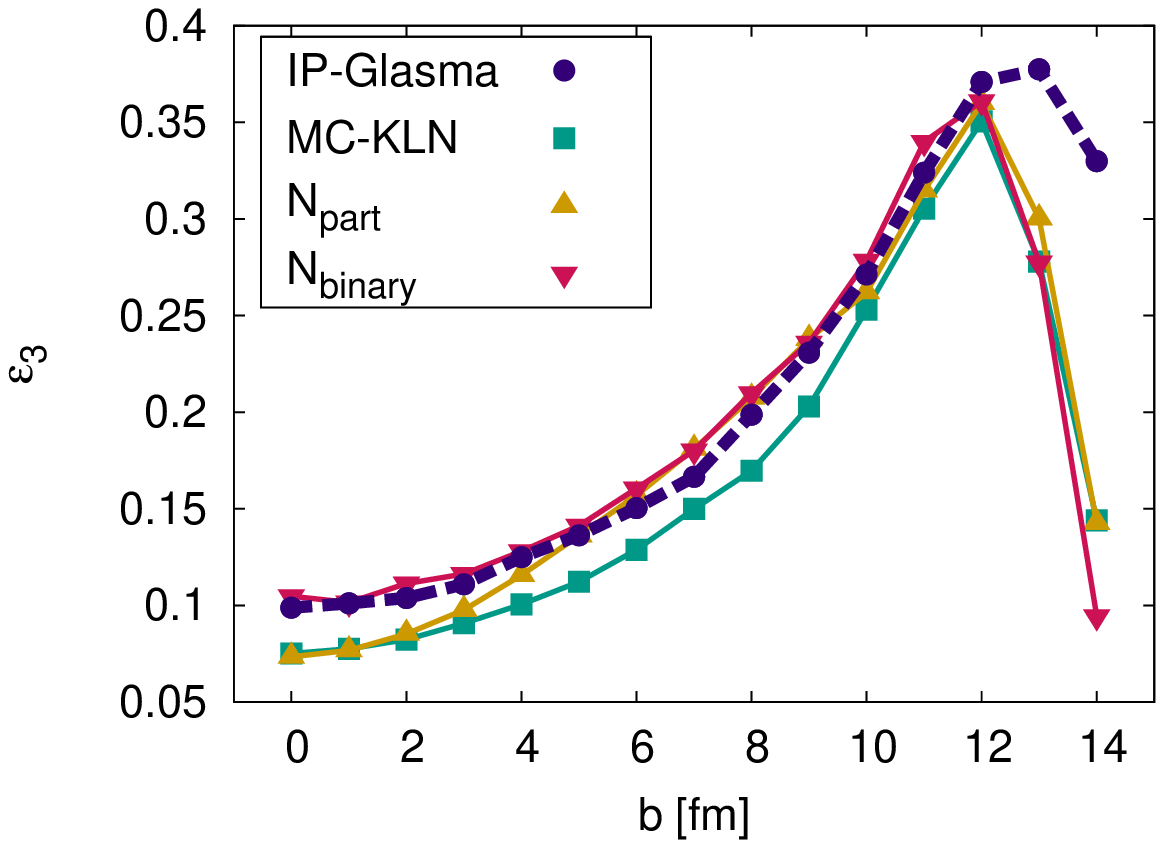}
     \vspace{-0.3cm}
     \caption{(Color online) Average participant ellipticity (upper panel) and triangularity (lower panel) of the initial state. 
       This calculation (circles), MC-KLN (squares), 
       Glauber implementation with participant and binary collision scaling (triangles).}
     \label{fig:ecc}
   \end{center}
\vspace{-0.7cm}
\end{figure}

We next determine the participant ellipticity $\varepsilon_2$ and triangularity $\varepsilon_3$ of all models. 
Final flow of hadrons $v_n$ is to good approximation proportional to the respective $\varepsilon_n$ \cite{Qin:2010pf}, which makes these eccentricities 
a good indicator of what to expect for $v_n$. We define
\begin{equation}
 \varepsilon_n = \frac{\sqrt{\langle r^n \cos(n\phi)\rangle^2+\langle r^n \sin(n\phi)\rangle^2}}{\langle r^n \rangle}\,,
\end{equation} 
where $\langle \cdot \rangle$ is the energy density weighted average. The results from averages over $\sim 600$
 events for each point plotted are shown in Fig.\,\ref{fig:ecc}. The ellipticity is largest in the MC-KLN model and smallest in the MC-Glauber model with participant scaling of the energy density ($N_{\rm part}$).
The result of the present calculation lies in between, agreeing well with the MC-Glauber model using binary collision scaling ($N_{\rm binary}$). We note however that this agreement is accidental; binary collision scaling of eccentricities, as shown explicitly in a previous work applying average CYM initial conditions \cite{Lappi:2006xc}, does not imply binary collision scaling of multiplicities.

The triangularities are very similar, with the MC-KLN result being below the other models for most impact parameters. Again, the present calculation is closest to the MC-Glauber model with binary collision scaling. There is no parameter dependence of eccentricities and triangularities in the IP-Glasma results shown in Fig.\,\ref{fig:ecc}. It is reassuring that both are close to those from the MC-Glauber model because the  latter is tuned to reproduce data even though it does not have dynamical QCD fluctuations.

We have checked that our results for $\varepsilon_2,\varepsilon_3$ are insensitive to the choice of the lattice spacing $a$, despite a logarithmic ultraviolet divergence of the energy density at $\tau=0$ \cite{Lappi:2006hq}. They are furthermore insensitive to the choice of $g$, the ratio $g^2\mu/Q_s$, and the uncertainty in Bjorken $x$ at a given energy.

\begin{figure}[tb]
   \begin{center}
     \includegraphics[width=8cm, clip, trim=0cm 0.35cm 0cm 0cm]{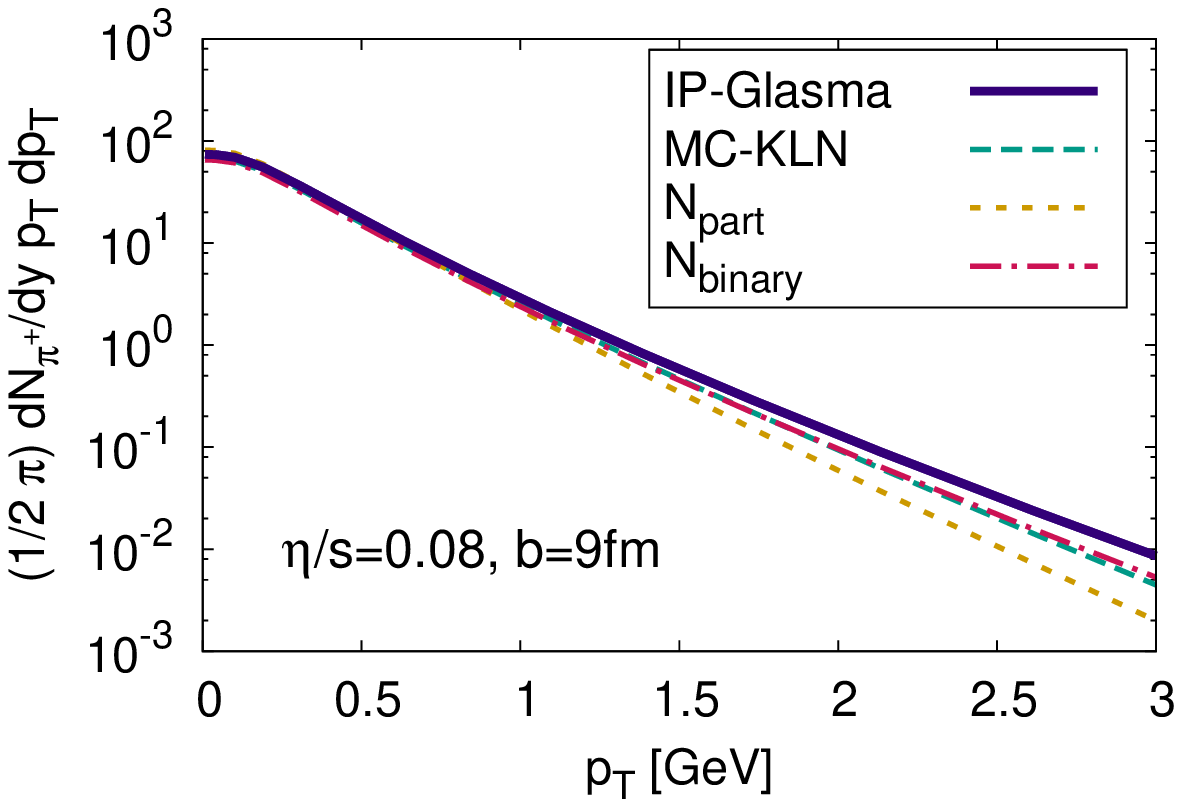}
     \includegraphics[width=8cm, clip, trim=0cm 0cm 0cm 0.4cm]{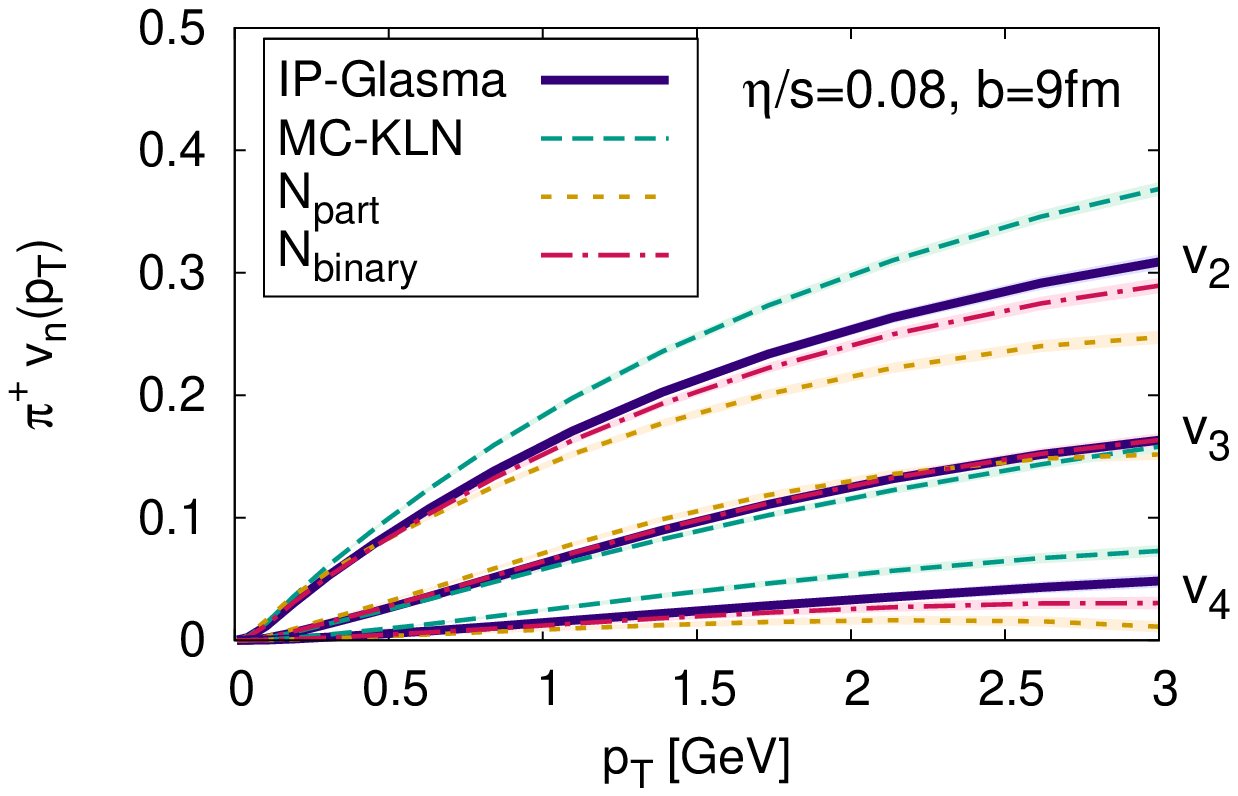}
     \vspace{-0.5cm}
     \caption{(Color online) Thermal $\pi^+$ transverse momentum spectra (upper) and anisotropic flow coefficients $v_2$, $v_3$, and $v_4$ 
         as functions of $p_T$ (lower) from IP-Glasma initial conditions (solid), MC-KLN (dashed),
       MC-Glauber using participant scaling (dotted) and binary collision scaling (dash-dotted). \label{fig:vn}}
   \end{center}
    \vspace{-1cm}
 \end{figure}

Finally, in Fig.\,\ref{fig:vn} we present results for the transverse momentum spectrum and anisotropic flow of thermal pions after evolution 
using \textsc{music} \cite{Schenke:2010nt,Schenke:2010rr} with boost-invariant initial conditions 
and shear viscosity to entropy density ratio $\eta/s=0.08$. 
Average maximal energy densities of all models were normalized to assure similar final multiplicities.
More pronounced hot spots, as emphasized previously~\cite{Gyulassy:1996br}, affect the particle spectra obtained from flow, leading to harder momentum spectra in the present calculation compared to MC-KLN and MC-Glauber models.
Differences in $v_2(p_T)$ and $v_3(p_T)$ are as expected from the initial eccentricities of the different models.

As discussed at the outset, MC-KLN fails to describe experimental $v_2$ and $v_3$ simultaneously \cite{Esumi:2011nd,Qiu:2011hf} 
because of its small ratio $\varepsilon_3/\varepsilon_2$.
The fluctuating IP-Glasma initial state presented here has a larger $\varepsilon_3/\varepsilon_2$, 
closer to that of the MC-Glauber model that is tuned to describe experimental $v_n$ reasonably well \cite{Schenke:2011bn}. 

In summary, we introduced the IP-Glasma model of fluctuating initial conditions for heavy-ion collisions.
This model goes beyond the MC-KLN implementation by using CYM solutions instead of  $k_\perp$-factorization and 
including quantum fluctuations on the dynamically generated transverse length scale $1/Q_s$.
Further, unlike MC-KLN, its parameters are fixed by HERA inclusive and diffractive e+p DIS data.
At fixed impact parameter, this model naturally produces NBD multiplicity fluctuations that are known to describe $p+p$ and $A+A$ multiplicity distributions, and its ratio of initial triangularity to eccentricity is more compatible with experimental data of harmonic flow coefficients.

Looking forward, an improved matching to the hydrodynamic description, starting at time $\tau_0$, can 
be achieved by including classical Yang-Mills evolution of the system up to this time. However, we do not expect a significant modification of the presented results for $\varepsilon_2$ and $\varepsilon_3$ as suggested by previous work~\cite{Lappi:2006xc}. Further refinements include treating color charge correlations encoded in the JIMWLK hierarchy for improved rapidity and energy distributions~\cite{Dumitru:2011vk,Iancu:2011nj} and eliminating arbitrariness in choice of thermalization time by an ab initio treatment of thermalization~\cite{Dusling:2011rz,Blaizot:2011xf,Kurkela:2011ub,Berges:2012iw}.  Detailed studies of higher flow harmonics using dissipative hydrodynamic simulations and comparison to experimental data will allow for further discrimination between different initial conditions. Specifically, it would be interesting to see if these comparisons are able to distinguish between our Glasma flux tube scenario with granularity on the energy dependent scale $1/\qs$ and other non-perturbative string scenarios which share common features such as NBD fluctuations but are sensitive to $1/\Lambda_{\rm QCD}$~\cite{Wang:1991hta,Magas:2000jx,*Magas:2002ge,Flensburg:2011kj}. 


\emph{Acknowledgments}
\hyphenation{Dumitru}
We thank S. Chattopadhyay, A. Dumitru, C. Gale, S. Jeon, T. Lappi, L. McLerran, and Z. Qiu for helpful discussions.
BPS\ and RV\ are supported by US Department of Energy under DOE Contract No.DE-AC02-98CH10886 and acknowledge additional support from a BNL ``Lab Directed Research and Development'' grant LDRD~10-043.

\vspace{-0.3cm}
\bibliography{spires}

\end{document}